\documentclass[aip,jcp,reprint]{revtex4-1} 
\usepackage{graphicx}

\usepackage{amsmath}    
\usepackage{graphicx}   
\usepackage{dcolumn}
\usepackage{color}
\usepackage{ulem}
\usepackage{caption}
\usepackage{subcaption}

\usepackage{array}
\usepackage{booktabs}
\usepackage{multirow}

\usepackage{booktabs}

\begin{document}
\preprint{AIP/123-QED}
\title{On the Role of Solvent in Hydrophobic Cavity-ligand Recognition Kinetics}
\author{Navjeet Ahalawat} \affiliation{Tata Institute of Fundamental Research, Center for Interdisciplinary sciences, Hyderabad 500107, India}\affiliation{Department of Molecular Biology, Biotechnology and Bioinformatics, Chaudhary Charan Singh Haryana Agricultural University, Hisar 125004, India} \author{Satyabrata Bandyopadhyay}\affiliation{Tata Institute of Fundamental Research, Center for Interdisciplinary sciences, Hyderabad 500107, India}\author{Jagannath Mondal} \email{jmondal@tifrh.res.in, +914020203091} \affiliation{Tata Institute of Fundamental Research, Center for Interdisciplinary sciences, Hyderabad 500107, India} 
\date{\today}

\begin{abstract}

Solvent often manifests itself as the key determinant of the kinetic aspect of molecular recognition process. While the solvent is often depicted as a source of barrier in the ligand recognition process by polar cavity,  the nature of solvent's role in the recognition process involving hydrophobic cavity and hydrophobic ligand remains to be addressed. In this work, we quantitatively assess the role of solvent in dictating the kinetic process of recognition in a popular system involving hydrophobic cavity and ligand. In this prototypical system the hydrophobic cavity undergoes \textit{dewetting transition} as the ligand approaches the cavity, which influences the cavity-ligand recognition kinetics. Here, we build  Markov state model (MSM) using adaptively sampled unrestrained molecular dynamics simulation trajectories to map the kinetic recognition process. The MSM-reconstructed free energy surface recovers a broad water distribution at an intermediate cavity-ligand separation, consistent with previous report of dewetting transition in this system. Time-structured independent component analysis of the simulated trajectories quantitatively shows that cavity-solvent density contributes considerably in an optimised reaction coordinate involving cavity-ligand separation and water occupancy.  Our approach quantifies two solvent-mediated macro states at an intermediate separation of the cavity-ligand recognition pathways, apart from the fully ligand-bound and fully ligand-unbound macro states. Interestingly, we find that these water-mediated intermediates, while transient in populations, can undergo slow mutual interconversion and create possibilities of multiple pathways of cavity recognition by the ligand. Overall, the work provides a quantitative assessment of the role that solvent plays in facilitating the recognition process involving hydrophobic cavity.

\end{abstract}

\maketitle

\newpage

\section{Introduction}
 
 Obtaining mechanistic insights into protein-ligand recognition process at an atomistic resolution has always remained a key topic of interest. Towards this end, over the last decade computer simulations\cite{shaw_dasatinib,dror_gpcr} have emerged as a major approach in deciphering the kinetic pathways leading to recognition of small molecules by the protein. By harnessing the computing powers obtained by innovations in hardwares and via smart usage of statistical model building approaches and enhanced sampling techniques, molecular dynamics simulations have come a long way in successfully elucidating the full kinetic process of ligand approaches in the deeply buried cavity in protein.\cite{mondal_t4l,mondal_P450,noe_trypsin,tiwary_dasatinib,defabritiis} While doing so, these simulations have been able to identify multiple key factors which influence the recognition process. 
 
 While in many cases, the subtle conformational fluctuations of the proteins (for example, transient opening of helix-gate\cite{mondal_t4l}, ligand-induced hydrophobic gate opening\cite{mondal_P450}) have been found to act as the trigger for the ligand recognition processes, solvent molecules have also been found to play a decisive factor in several cases of recognition process. Especially, seminal simulations by  Shaw and coworkers have reported that kinase inhibitors\cite{shaw_dasatinib} and drugs targeting GPCRs\cite{dror_gpcr} need to displace certain solvent molecules, before accessing the binding cavity. In related findings, Mondal et al\cite{mondal_dasatinib} have estimated the extent of desolvation barriers that the kinase inhibitors need to overcome  prior to ligand access to the protein cavity. Tiwary et al\cite{tiwary_dasatinib} have reported coupled protein-water movements through multiple metastable intermediates that orchastrate exit process of kinase inhibitor from the binding site.  In special scenario involving hydrophobic cavity and ligand, solvent molecules (water) have been found to undergo \textit{dewetting transitions} \cite{chandler,berneweekszhou,hummer} in the hydrophobic cavity, which, in turn, influences the kinetics of ligand recognition process\cite{young2007motifs,hiv}. The current work provides a quantitative account of kinetics underlying hydrophobic cavity-ligand recognition in a popular system\cite{mondal_fuller, tiwary2015} manifesting significant dewetting transition.    
 
 The key system of interest in the current work is a fullerene molecule recognising a non-polar uncharged concave cavity solvated in water (Figure ~\ref{snap}). Over the last several years, this prototypical systems\cite{mondal_fuller, tiwary2015,tiwary_wet} and related systems\cite{setny_jctc,mccammon_jacs, setny_pnas,setny_jcp} have served as a model of numerous past computer simulations and theoretical studies investigating the role of water in the molecular recognitions. As an important finding involving this system, Mondal, Morrone and Berne had earlier shown\cite{mondal_fuller} that  a moderately hydrophobic cavity is capable of demonstrating a classic case of \textit{dewetting transition}: at an intermediate cavity-ligand separation, the cavity undergoes periodic cycle of water molecules replenishing the cavity and receding from it.  As a significant outcome of this dewetting transition, the water fluctuation in the cavity was found to be correlated with  position-dependent ligand diffusion:  the ligand diffusion coefficient corresponding to dewetting transition at intermediate cavity-ligand separation was estimated to be slowest. The work further showed that this position dependent ligand diffusion due to the dewetting transition ultimately influences the kinetics of ligand recognition process to the cavity.  Overall, the investigation , albeit qualitatively, hypothesised that water occupancy, apart from the cavity-ligand separation, needs to be explicitly taken into account, to justify the slow ligand recognition kinetics into the cavity. A subsequent work by Tiwary, Mondal, Morrone and Berne explored\cite{tiwary2015} the thermodynamics and kinetics of the unbinding process of the fullerene from the same cavity using meta dynamics simulations and have reinforced the original hypothesis of water's role in controlling the recognition process, especially when the ligand is restricted to move along axial direction of the cavity. 
 
 This current work aims at using unbiased Molecular Dynamics simulations and Markov state model to quantify the extent of water-mediation in driving the hydrophobic cavity-ligand recognition process. In this article, we specifically ask following questions: 
 \begin{itemize}
 \item What is the relative contribution of water occupancy in describing the recognition process?
 \item Can we characterise the water-mediated intermediates involving cavity-ligand recognition?
 \item What are the kinetic pathways of cavity-ligand recognition and what role do the intermediates play in driving the recognition process? 
  \end{itemize}
  
  Towards this end, first  we reconstruct the free energy surfaces along cavity-ligand separation($q$) and water-occupancy($N_w$). Subsequently we quantify the relative contribution of cavity water occupancy in an optimized collective variable as the linear combination of $q$ and $N_w$, within the framework of time structured independent component analysis (TICA)\cite{tica1,tica5}. Second, we quantitatively map the complete recognition process by combining numerous adaptively sampled short unbiased MD simulation trajectories of cavity-ligand approaches in solvent with a comprehensive Markov state model (MSM)\cite{msm_review}, characterise the stationary population of key macro states and subsequently connect them via transition path theory. As an important finding, we discover that the water-mediated intermediates, while transient in nature, give rise to multiple pathways of ligand access to the hydrophobic cavity, an observation reminiscent of our recent report\cite{mondal_t4l} of multiple pathways in small molecular recognising the cavity of protein L99A T4 Lysozyme.

\section{Material and Methodology}

\textit{System preparation and molecular dynamics simulation protocol}

\begin{figure}[htp]
\includegraphics[clip,width=1.0\columnwidth,keepaspectratio]{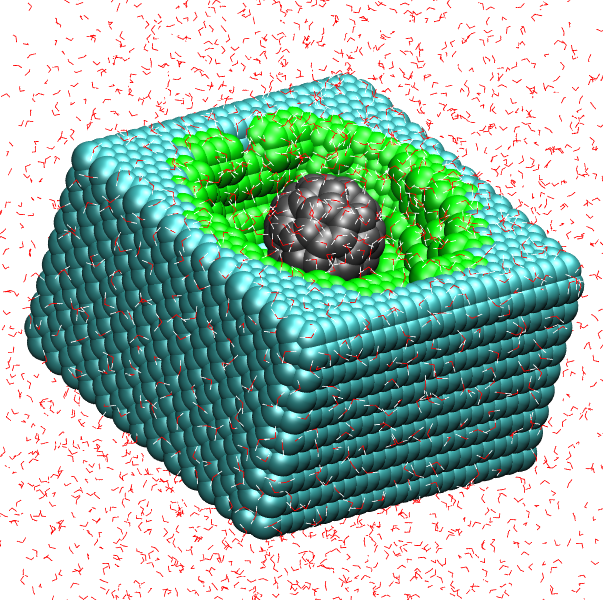}
\caption{Representative snapshot of hydrophobic cavity and spherical ligand bound to it. The hexagonal closed packed wall atoms are shaded with cyan color, while the cavity atoms are shaded with green color. Fullerene molecule is modelled as a spherical ligand and is shaded with black color. The model is identical to that used earlier by Berne and coworkers\cite{mondal_fuller,tiwary2015}.}
\label{snap}
\end{figure}

The system of interest in this work involves an uncharged pocket with a concave ellipsoidal hole carved from hydrophobic slab and a fullerene molecule as the ligand. Both of them are solvated by water molecules. A representative configuration of the system is depicted in figure ~\ref{snap}.  Similar to previous works,\cite{mondal_fuller,tiwary2015} we consider the attraction between water and both the pocket and ligands weak and almost mimicking a hydrophobic system. The pocket sites comprise of two types of atomic species: i) cavity atoms (coloured green in Figure ~\ref{snap}) with Lenard Jones (LJ) parameter $\epsilon$= 0.008 kJ/mol and $\sigma$=0.4152 nm ii) wall atoms (coloured cyan in Figure ~\ref{snap}) with LJ parameter $\epsilon$= 0.0024 kJ/mol and $\sigma$=0.4152 nm. Fullerene atoms are modelled by LJ parameter $\epsilon$= 0.276 kJ/mol and $\sigma$=0.350 nm. All simulations are performed in explicit water described by TIP4P model.\cite{tip4p} The solute-solvent interactions were modelled by geometric combinations of solvent and solute parameters. The simulation parameters are identical to that previously used by Mondal, Morrone and Berne\cite{mondal_fuller} and Tiwary, Mondal, Morrone and Berne\cite{tiwary2015}.

In the current work, we initiate a large set of unbiased molecular dynamics simulations at diverse cavity-ligand separations with randomly distributed velocity seeds. Each of the simulations are 500 picosecond long and an aggregate of 4048 copies of simulations have been performed.  In all our simulations, the cavity is kept fixed and  the ligand is statically constrained to move along axis of symmetry of cavity and ligand. We subsequently sample the movements of ligand and solvent molecules by molecular dynamics (MD) simulations. We emphasise that, as a key departure from earlier simulation protocol of Mondal, Morrone and Berne\cite{mondal_fuller}, in this work we have not used any restraint for averting ligand diffusion in the bulk solvent.  All simulations are performed in NVT ensemble with the temperature being maintained at 300 K by stochastic velocity rescaling thermostat.\cite{bussidonadio} The dimension of the simulation box was 4.55 nm $\times$ 4.55 nm $\times$ 5.93 nm. The system comprised of a total of 22608 particles, with the total number of ligand, cavity and solvent atoms equalling 60, 9020 and 13528 atoms (3382 TIP4P water molecules) respectively. An integration time-step of 2 femto second was  used for all MD simulations. Particle Mesh Ewald summation technique were employed for treatment of long-range interactions\cite{pme1}. Each of the simulation trajectories was analyzed to extract the time profile of cavity-ligand distance ($q$) and number of water molecules in cavity ($N_w$). Figure ~\ref{tica_fes} A schematically illustrate these two collective variables.

\textit{Development of Markov state model}

\begin{figure}[htp]
\includegraphics[clip,width=1.0\columnwidth,keepaspectratio]{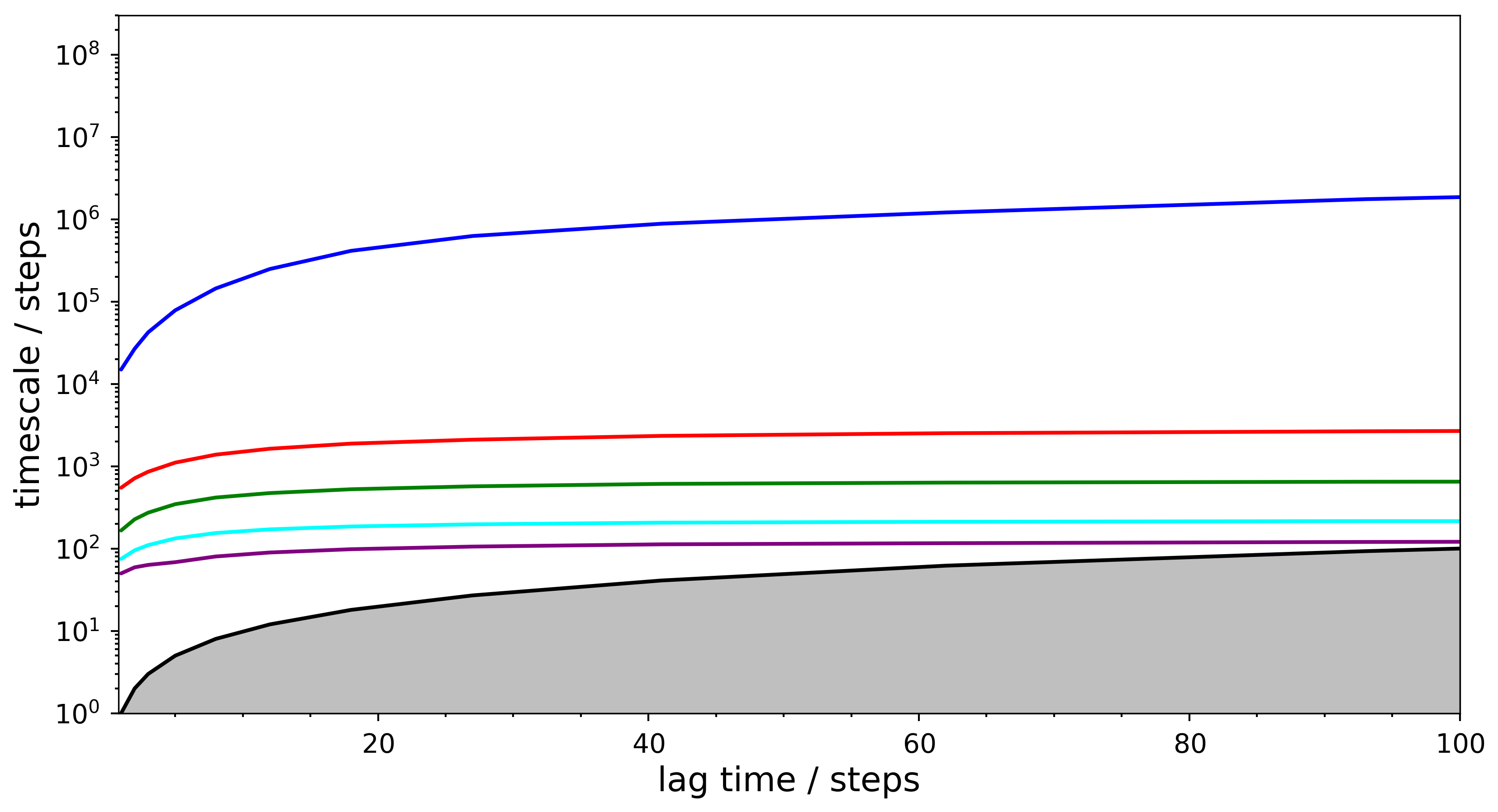}
\caption{Implied time scale plot of underlying Markov state model. A lag time of 20 time step, which corresponds to 10 ps, was considered as the lag time for building MSM for the present work. }
\label{its}
\end{figure}

The series of MD simulation trajectories generated in this work were eventually used to build a Markov state model (MSM)\cite{msm_review} for statistically mapping the complete process of cavity-ligand recognition.We employed Pyemma software for this purpose\cite{pyemma,pyemma_tutorial}. Towards this end, continuous trajectories were first discretised by clustering the trajectories  using cavity-ligand separation and water numbers in the cavity as the metric. K-means clustering approach\cite{kmeans} was used for this purpose and 500 cluster centres were used to discretize the data. Subsequently, the discretised data set is then built into a Markov state model.  A lag time of 10 picosecond (equivalent to 20 time step) was chosen for building the MSM, as the implied time scale of the MSM was found to reach plateue beyond this. (See figure ~\ref{its}) The stationary populations of the discrete microstates were computed from the MSM. These stationary populations were used to reweigh the free energy surfaces derived  from these short non-equilibrium trajectories.\cite{pyemma,pyemma_tutorial,pyemma_github} 

The time structured independent component analysis (TICA)\cite{tica1,tica5} is employed to quantify the relative contribution of solvent in an optimised collective variable for describing the cavity-ligand recognition process presented in this work. As has been originally formulated by Pande and coworkers\cite{tica5} and Noe and coworkers\cite{tica1} and as has been recently utilised by us in a recent article\cite{mondal_gb1}, TICA maximises the autocorrelation of projection of a collective variable and deduce the slowest collective variable as the linear combination of the constituent collective variables. In this work, we employ TICA  to express two hybrid collective variables ($TIC_1$ and $TIC_2$), for describing the cavity-ligand recognition process, as the linear combinations of two collective variables namely, cavity-ligand distance ($q$) and the number of waters in the cavity($N_w$) ( see Figure ~\ref{tica_fes} A for schematic illustration) .

\begin{equation}
TIC_1 = c_{q}\widetilde{\phi}_{q} + c_{N_w}\widetilde{\phi}_{N_w}   \\
\label{eq:pref_bind2}
\end{equation}

\begin{equation}
TIC_2 = {c}'_{q}\widetilde{\phi}_{q} + {c}'_{N_w}\widetilde{\phi}_{N_w}  \\
\end{equation}

where $\widetilde\phi_{k}$ represents k-th mean free input CV and $c_k$ and ${c}'_k$ are associated coefficients for $TIC_1$ and $TIC_2$ respectively.  All the input CVs were first rescaled between 0 and 1 before using them to build the TICA model.
The modulus of the coefficients in the linear combinations provide relative contribution of each of the constituent collective variables in the optimised one.

One of the main purposes of the manuscript is to quantify the kinetics of transition among the key pocket-ligand configurational states described by Mondal, Morrone and Berne\cite{mondal_fuller}. Towards this end, we lumped the relevant micro states into four macro states, based on the nature of free energy surfaces and prior description by Mondal, Morrone and Berne:  bound [$q \leq1.0$ nm], dry intermediate ($I_{dry}$) [$1.0<q \leq1.4$ nm and $N_w \leq16$], wet intermediate ($I_{wet}$) [$1.0<q \leq1.4$ nm and $N_w>16$],  and unbound [$q>1.4$ nm]. Subsequently, transition path theory (TPT)\cite{tpt1,tpt3} was employed using Pyemma software\cite{pyemma} to deduce the kinetic networks and mean first passage time (MFPT) among the macro states and their relative contributions  in the paths leading to the cavity-ligand recognitions.

\section{Results and Discussion}

\begin{figure}[htp]
\includegraphics[clip,width=1.0\columnwidth,keepaspectratio]{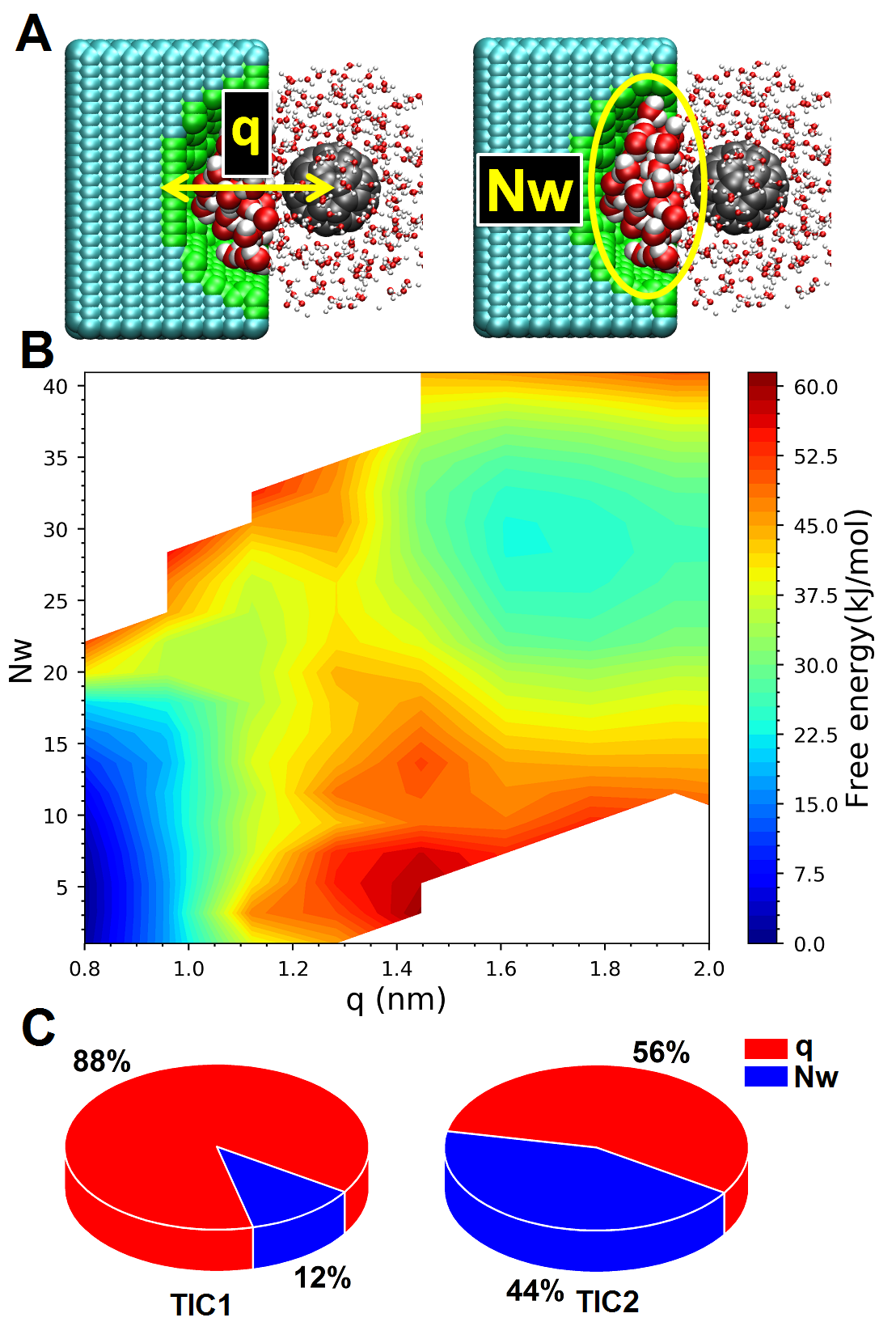}
\caption{A. Schematic illustration of two collective variables of interest ($N_w$ and $q$). B. Free energy surface after reweighing the trajectories by stationary populations of the MSM-generated microstates. C. Contribution of different collective variable along TICA. }
\label{tica_fes}
\end{figure}

\begin{figure}[htp]
\includegraphics[clip,width=1.0\columnwidth,keepaspectratio]{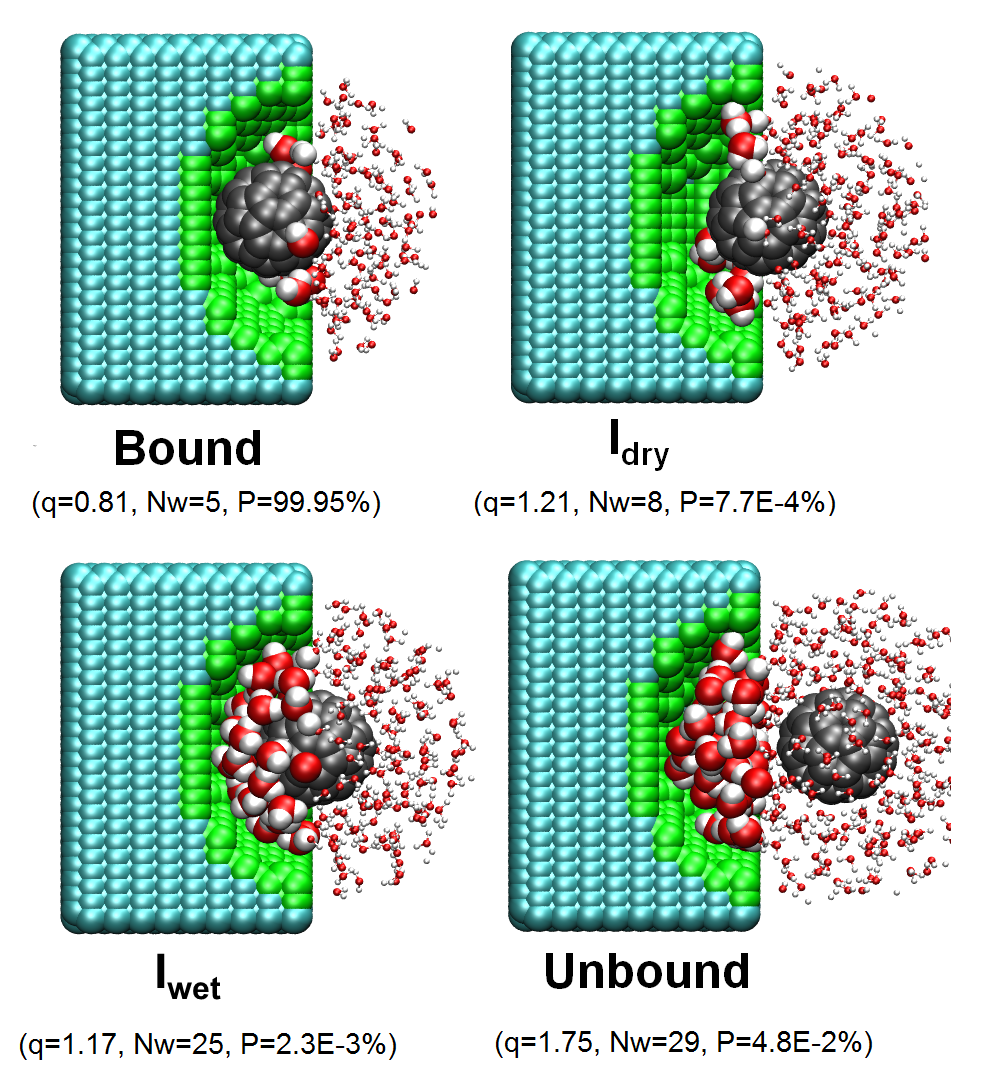}
\caption{Representative snapshots of different macrostates obtained after clustering are shown with the details of a stationary population of macro state, distance of ligand from the cavity ($q$), and the number of water molecules ($N_w$) in that snapshot.}
\label{state}
\end{figure}

A major observation from the previous studies\cite{mondal_fuller,tiwary2015} on the current system of interest had been that the solvent (water) plays a crucial role in describing the overall cavity-ligand recognition process in this system. In a bid to understand the underlying thermodynamic landscape, we first map the free energetics of the recognition process. The two-dimensional free energy surface along cavity-ligand distance ($q$) and water-occupancy inside the cavity ($N_w$), derived from aggregated short trajectories and reweighed by the MSM-derived stationary populations of the discrete micro-states,\cite{pyemma_tutorial,pyemma_github} is shown in figure ~\ref{tica_fes} B. We find that the free energy surface correctly predicts the `dry' ligand-bound state as the free energy minima ($q$=0.81 nm, $N_w$=5). (See representative snapshot in figure ~\ref{state}, top left) On the other hand, the free energy increases with increase in cavity-ligand distance and reaches a plateau at `wet' ligand-unbound state corresponding to $q$=1.75 nm and $N_w$=29. (see representative snapshot in figure ~\ref{state}, Bottom, right) The overall free energy profile, here obtained via combination of unbiased simulations and MSM approach, is in semi-quantitative agreement with earlier reported free energetics via meta dynamics simulation approaches by Tiwary, Mondal, Morrone and Berne\cite{tiwary2015}. The current work predicts a slightly lower free energy barrier ($\approx$ 45 kJ/mol) at an intermediate cavity-ligand separation ($q$ $\approx$ 1.4 nm)  than the earlier report of Tiwary, Mondal, Morrone and Berne ($\approx$ 60 kJ/mol). On the other hand, net binding free energy estimated by the current free energy surface (30 kJ/mol) is relatively lower than the one estimated by earlier report (60 kJ/mol) based on metadynamics simulation. A possible reason for the relatively lower free energy of binding predicted by the current estimate than the earlier work is that, unlike the previous works,  this work does not employ any restraining potential in an intermediate cavity-ligand separation, thereby allowing ligand diffusion in the bulk.

As described in the introduction of the article, one of the key goals of the current work is to quantify the relative contributions of water occupancy inside the cavity ($N_w$) and pocket-ligand distance ($q$) in an optimised collective variable involving these two constituent collective variables (see figure ~\ref{tica_fes} A for illustration of these two collective variables.) Towards this end, we express the optimised collective variables ($TIC_1$ and $TIC_2$)  as a linear combination of $N_w$ and $q$ within the frame-work of time-structured independent component analysis (TICA) . $TIC_1$ and $TIC_2$ are respectively slowest and second slowest collective variables (see equation 1 and 2). The modulus of the coefficients in $TIC_1$ and $TIC_2$ provides relative weights of the $N_w$ and $q$ in the optimised collective variables. We find that at the lag time of 10 ps, at which the MSM is constructed, the relative contribution of $N_w$ in $TIC_1$ is 12 \%. However, along $TIC_2$ the relative contribution of $N_w$ is more significant (44 \%). (See figure ~\ref{tica_fes}C) For this same system, Mondal, Morrone and Berne\cite{mondal_fuller} have previously demonstrated that the cavity undergoes considerable solvent density fluctuations with a bimodal water distribution at a critical intermediate cavity-ligand separation. This earlier observation justifies non-zero contribution of $N_w$ in the optimally constructed collective variables in the current work via TICA approach. This result is also consistent with earlier report of a 'wet' optimal reaction coordinate for this current system by Tiwary and Berne\cite{tiwary_wet}, which predicted a 7.5 \% contribution of cavity-water occupancy in the overall optimised reaction coordinate, albeit using spectral gap optimization approach\cite{sgoop}. Overall, our analysis suggests that an optimal reaction coordinate describing the cavity-ligand recognition process would have significant contribution from cavity-water density,  apart from the more obvious metric namely cavity-ligand separations.

 Interestingly, our MSM-derived free energy surface ( figure ~\ref{tica_fes} B) recovers a broad water distribution corresponding to $q$ $\approx$ 1.2 nm, which had been previously identified by Mondal, Morrone and Berne\cite{mondal_fuller} as the region of \textit {dewetting transition}.  However, we also note that while the current free energy surface (figure ~\ref{tica_fes}) shows a local minima at around $q$ $\approx$ 1.12 nm for high water content ($N_w$=25), corresponding to 'wet intermediate' ($I_{wet}$) (see representative snapshot in figure ~\ref{state}, bottom, right) and a broad water distribution around the same cavity-ligand distance, it does not show a clear local basin for `dry intermediate' at a very weak water content corresponding to similar $q$=1.2 nm. Nonetheless, a major hypothesis of previous work by Mondal, Morrone and  Berne\cite{mondal_fuller} was that the recognition of the fullerene into the cavity is orchestrated by two-dimensional hopping between wet and dry states and `dewetting transition' at an intermediate pocket-ligand separation. These earlier observations prompted us to assess the role of these intermediates in the cavity-ligand recognition process.  Towards this end, we discretized all simulated trajectories using $N_w$ and $q$ as the clustering metric and built a Markov state model (MSM) on top of the free energy surface(see methods). The micro states constituting the MSM were subsequently coarse-grained into four relevant cavity-ligand macro states, in accordance with the state descriptions by early work of Mondal, Morrone and Berne\cite{mondal_fuller}. Figure ~\ref{state} depicts the representative snapshots of four relevant cavity-ligand macro states: `bound' state with $q$=0.81 nm and $N_w$=5, `unbound' state with $q$=1.75 nm and $N_w$=29, `dry intermediate' ($I_{dry}$)' with $q$=1.21 nm and $N_w$=8 and `wet intermediate($I_{wet}$)' with $q$=1.17 nm and $N_w$=25. We also derive the stationary population of each of the macro states from MSM. We find that majority of the equilibrium population involves 'bound' state (99.95 percent) while `unbound' state has tiny stationary population of 0.048 percent, consistent with the presence of deep free energy minima corresponding to the 'bound' macrostate in the free energy surface (figure ~\ref{tica_fes} B). On the other hand the `wet intermediate ($I_{wet}$)' and especially 'dry intermediate ($I_{dry}$)' appears to be transiently populated with minor populations of 2.3 X $10^{-3}$ and $7.7 X 10^{-4}$ percent respectively.

\begin{figure}[htp]
\includegraphics[clip,width=1.0\columnwidth,keepaspectratio]{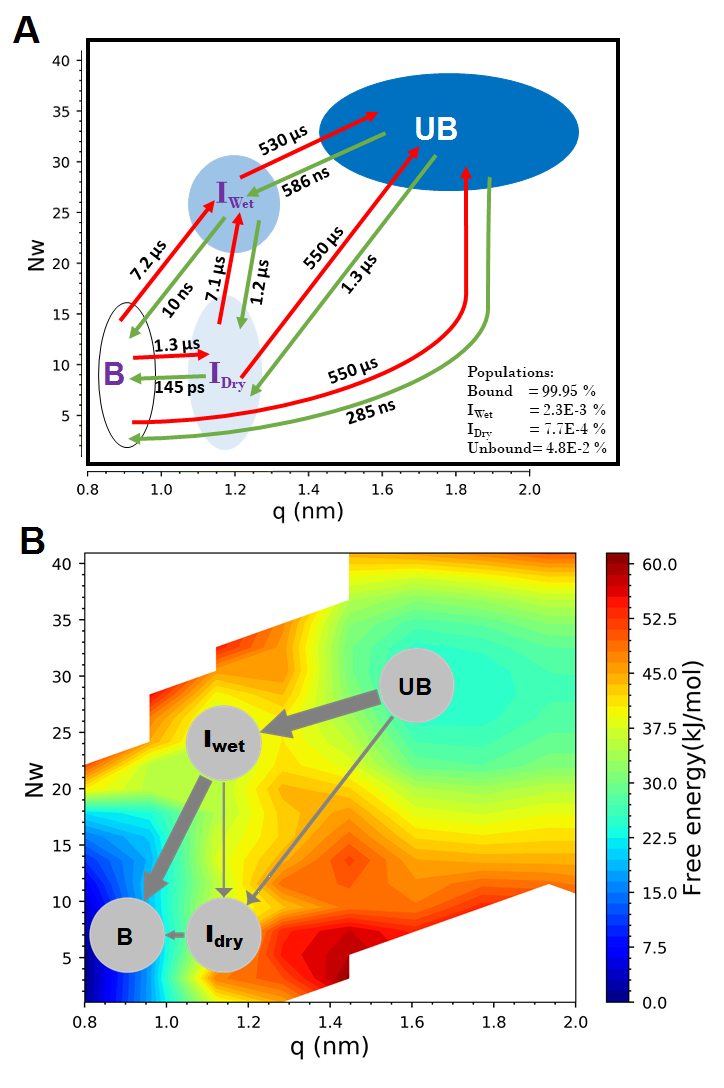}
\caption{A. Schematic representation of different states of the cavity-ligand system with respect to q and $N_w$, where elliptical shapes are macrostates and the transition time (MFPT) from one state to another state is shown by arrow. B.  Transition pathways from unbound (UB) state to bound (B) state shown on the top of free energy surface where the circle represents the macrostate and arrow thickness represents the net flux going from one state to another state. Relative contribution of pathways: UB $\rightarrow$ $I_{wet}$ $\rightarrow$ 'B' 73.4 \%, UB $\rightarrow$ $I_{dry}$ $\rightarrow$ 'B' 18.1 \%, UB$\rightarrow$ $I_{wet}$ $\rightarrow$ $I_{dry}$ $\rightarrow$ B 8.5 \%  }
\label{mfpt_tpt}
\end{figure}

A key observation of the previous reports of Mondal, Morrone and Berne\cite{mondal_fuller} was large fluctuation in water-occupancy at the intermediate pocket-ligand distance. In the present work, we wanted to understand the time-scale of the kinetic interconversion among the above mentioned macro states. Towards this end, we apply transition path theory (TPT)\cite{tpt1,tpt3,pyemma} on the four macro-states and estimate the state-to-state mean first passage time (MFPT) (shown in figure ~\ref{mfpt_tpt} A). Especially the MFPT for interconversion between the two intermediate macro states $I_{dry}$ and $I_{wet}$ provides key insights about the kinetic reversible interconversion among the above mentioned macro states. We note that any transition leading from the `$I_{dry}$' state $\rightarrow$ `$I_{wet}$' state is about six times slower (MFPT= 7.1 micrsecond) the reverse transition (MFPT for `$I_{wet}$' state $\rightarrow$ `$I_{dry}$' = 1.2 microsecond). Overall, our assessment  suggests that while  these two water-mediated intermediates can undergo slow interconversion, the interconversion among these two intermediates is not perfectly reversible, with the equilibrium shifting more towards the dry state.   MFPT for Unbound $\rightarrow$ bound macro state interconversion and the reverse process (Bound $\rightarrow$ unbound interconversion) respectively  shed light on the `on' and `off' process of the cavity-ligand recognition process. As represented in figure ~\ref{mfpt_tpt}, our approach  predicts a binding time of 285 nanosecond and unbinding time of 550 microsecond for the cavity-ligand recognition. Our estimate of binding MFPT (285 nanosecond) based on unbiased simulation trajectories and MSM analysis is relatively slower than predicted earlier (769 picosecond) by Tiwary, Mondal, Morrone and Berne\cite{tiwary2015} based on the metadynamics simulations. On the other hand, the current estimate of unbinding MFPT (550 microsecond), as inferred from the Markov state modelling of the unbiased trajectories, is relatively faster than that of Tiwary, Mondal, Morrone and Berne (3863 second)\cite{tiwary2015} as estimated from infrequent metadynamics simulation. We believe that a possible cause of current work's estimates of relatively slower ligand-binding time and relatively faster ligand-unbinding time compared to that of previous works is to do with the fact the current work does not enforce any repulsive wall at an intermediate cavity-ligand separation, unlike the previous works\cite{mondal_fuller,tiwary2015}. As a result of the lack of the repulsive wall, the ligand entry towards the pocket is found to be slower and ligand exit away from the pocket is found to be faster than the earlier reports.

Given transient stationary populations of these water-mediated intermediate states, a pertinent question is whether these transient intermediates might play any role in deciding the binding kinetics of the ligand to the cavity. To address this question, we analyse possible flux leading from `unbound' macro state to `bound' macro state, within the framework of transition path theory.  As shown in figure ~\ref{mfpt_tpt}B, the transition paths leading from `unbound' macro state to 'bound' macro state provide key insights into possible paths of ligand recognition and respective contributions of these paths in ligand recognition. Firstly, we find that the ligand adopts a mechanism of \textit{multiple pathways} in its bid for transition from 'unbound' state to the `bound' state. Secondly and most intriguingly, we find that the pathways via `wet intermediate ($I_{wet}$)' or 'dry intermediate ($I_{dry}$)' contribute an aggregate of around 90 percentage among all pathways. Specifically, Unbound $\rightarrow$ $I_{wet}$ $\rightarrow$ `bound' pathway contributes 73.4 \% of the total path, while Unbound $\rightarrow$ $I_{dry}$ $\rightarrow$ 'bound' path contributes 18.1 \% of the total path. These two pathways are also in line with our finding of considerable contribution of solvent-occupancy in the overall optimised reaction coordinate.  More over, as shown in figure ~\ref{mfpt_tpt}B, we recover an additional path with 8.5 \% contribution where the ligand-cavity recognition can occur via Unbound $\rightarrow$ 'wet intermediate'($I_{wet}$) $\rightarrow$ 'dry intermediate' ($I_{dry}$) $\rightarrow$ 'bound' pathway.   Overall, this analysis of pathways clearly indicates that the intermediates, while transient, can act as the crucial mediators of  the ligand binding process.

\section{Conclusions}

In summary, the current article investigates the kinetic process of ligand recognition by a hydrophobic cavity. This prototypical system has in past served as a key model for multiple prior theoretical and simulation studies\cite{mondal_fuller,tiwary2015}. In this work, we explore a quantitative approach to revisit the potential role of solvent molecules in deciding the kinetic process of hydrophobic cavity-ligand recognition. We make use of TICA and quantify that cavity-water occupancy has significant contribution in an effective reaction coordinate describing the recognition process. This observation supports the previous report of 'wet' reaction coordinate for this system\cite{tiwary_wet}, especially when the ligand is statically constrained to move along the axis of symmetry.  Subsequently we map the network of recognition process of ligand by the cavity by combining numerous short Molecular Dynamics simulation with MSM, characterise the stationary population of key macro states and finally connect them via transition path theory. The MSM-reconstructed free energy surface along cavity-ligand separation and distance identifies a broad water distribution at an intermediate cavity-ligand separation, in consistence with previous report of large water fluctuation at similar separation. We quantify two water-mediated intermediates, apart from complete ligand-unbound and complete ligand-bound macro states. We find that while these two intermediates are transient in populations, they can undergo slow mutual interconversion and more interestingly create multiple pathways for ligand recognitions.  These observations of multiple pathways via short-lived intermediates are reminiscent of a recently investigated realistic protein-ligand recognition event\cite{mondal_t4l} involving L99A T4 Lysozyme and benzene molecule, where subtle helix-fluctuations of protein were found to give rise to a  pair of intermediates which dictate multiple pathways of protein-ligand recognition process.

\section*{Acknowledgments}
The current work was enabled by computing facilities developed using intramural research grants received from TIFR, India, Ramanujan Fellowship and Early Career Research funds provided by the Department of Science and Technology (DST) of India (ECR/2016/000672).


\begin{thebibliography}{34}%
\makeatletter
\providecommand \@ifxundefined [1]{%
 \@ifx{#1\undefined}
}%
\providecommand \@ifnum [1]{%
 \ifnum #1\expandafter \@firstoftwo
 \else \expandafter \@secondoftwo
 \fi
}%
\providecommand \@ifx [1]{%
 \ifx #1\expandafter \@firstoftwo
 \else \expandafter \@secondoftwo
 \fi
}%
\providecommand \natexlab [1]{#1}%
\providecommand \enquote  [1]{``#1''}%
\providecommand \bibnamefont  [1]{#1}%
\providecommand \bibfnamefont [1]{#1}%
\providecommand \citenamefont [1]{#1}%
\providecommand \href@noop [0]{\@secondoftwo}%
\providecommand \href [0]{\begingroup \@sanitize@url \@href}%
\providecommand \@href[1]{\@@startlink{#1}\@@href}%
\providecommand \@@href[1]{\endgroup#1\@@endlink}%
\providecommand \@sanitize@url [0]{\catcode `\\12\catcode `\$12\catcode
  `\&12\catcode `\#12\catcode `\^12\catcode `\_12\catcode `\%12\relax}%
\providecommand \@@startlink[1]{}%
\providecommand \@@endlink[0]{}%
\providecommand \url  [0]{\begingroup\@sanitize@url \@url }%
\providecommand \@url [1]{\endgroup\@href {#1}{\urlprefix }}%
\providecommand \urlprefix  [0]{URL }%
\providecommand \Eprint [0]{\href }%
\providecommand \doibase [0]{http://dx.doi.org/}%
\providecommand \selectlanguage [0]{\@gobble}%
\providecommand \bibinfo  [0]{\@secondoftwo}%
\providecommand \bibfield  [0]{\@secondoftwo}%
\providecommand \translation [1]{[#1]}%
\providecommand \BibitemOpen [0]{}%
\providecommand \bibitemStop [0]{}%
\providecommand \bibitemNoStop [0]{.\EOS\space}%
\providecommand \EOS [0]{\spacefactor3000\relax}%
\providecommand \BibitemShut  [1]{\csname bibitem#1\endcsname}%
\let\auto@bib@innerbib\@empty
\bibitem [{\citenamefont {Shan}\ \emph {et~al.}(2011)\citenamefont {Shan},
  \citenamefont {Kim}, \citenamefont {Eastwood}, \citenamefont {Dror},
  \citenamefont {Seeliger},\ and\ \citenamefont {Shaw}}]{shaw_dasatinib}%
  \BibitemOpen
  \bibfield  {author} {\bibinfo {author} {\bibfnamefont {Y.}~\bibnamefont
  {Shan}}, \bibinfo {author} {\bibfnamefont {E.~T.}\ \bibnamefont {Kim}},
  \bibinfo {author} {\bibfnamefont {M.~P.}\ \bibnamefont {Eastwood}}, \bibinfo
  {author} {\bibfnamefont {R.~O.}\ \bibnamefont {Dror}}, \bibinfo {author}
  {\bibfnamefont {M.~A.}\ \bibnamefont {Seeliger}}, \ and\ \bibinfo {author}
  {\bibfnamefont {D.~E.}\ \bibnamefont {Shaw}},\ }\href {\doibase
  10.1021/ja202726y} {\bibfield  {journal} {\bibinfo  {journal} {Journal of the
  American Chemical Society}\ }\textbf {\bibinfo {volume} {133}},\ \bibinfo
  {pages} {9181} (\bibinfo {year} {2011})},\ \bibinfo {note} {pMID: 21545110},\
  \Eprint {http://arxiv.org/abs/http://dx.doi.org/10.1021/ja202726y}
  {http://dx.doi.org/10.1021/ja202726y} \BibitemShut {NoStop}%
\bibitem [{\citenamefont {Dror}\ \emph {et~al.}(2011)\citenamefont {Dror},
  \citenamefont {Pan}, \citenamefont {Arlow}, \citenamefont {Borhani},
  \citenamefont {Maragakis}, \citenamefont {Shan}, \citenamefont {Xu},\ and\
  \citenamefont {Shaw}}]{dror_gpcr}%
  \BibitemOpen
  \bibfield  {author} {\bibinfo {author} {\bibfnamefont {R.~O.}\ \bibnamefont
  {Dror}}, \bibinfo {author} {\bibfnamefont {A.~C.}\ \bibnamefont {Pan}},
  \bibinfo {author} {\bibfnamefont {D.~H.}\ \bibnamefont {Arlow}}, \bibinfo
  {author} {\bibfnamefont {D.~W.}\ \bibnamefont {Borhani}}, \bibinfo {author}
  {\bibfnamefont {P.}~\bibnamefont {Maragakis}}, \bibinfo {author}
  {\bibfnamefont {Y.}~\bibnamefont {Shan}}, \bibinfo {author} {\bibfnamefont
  {H.}~\bibnamefont {Xu}}, \ and\ \bibinfo {author} {\bibfnamefont {D.~E.}\
  \bibnamefont {Shaw}},\ }\href {\doibase 10.1073/pnas.1104614108} {\bibfield
  {journal} {\bibinfo  {journal} {Proceedings of the National Academy of
  Sciences}\ }\textbf {\bibinfo {volume} {108}},\ \bibinfo {pages} {13118}
  (\bibinfo {year} {2011})},\ \Eprint
  {http://arxiv.org/abs/https://www.pnas.org/content/108/32/13118.full.pdf}
  {https://www.pnas.org/content/108/32/13118.full.pdf} \BibitemShut {NoStop}%
\bibitem [{\citenamefont {Mondal}\ \emph {et~al.}(2018)\citenamefont {Mondal},
  \citenamefont {Ahalawat}, \citenamefont {Pandit}, \citenamefont {Kay},\ and\
  \citenamefont {Vallurupalli}}]{mondal_t4l}%
  \BibitemOpen
  \bibfield  {author} {\bibinfo {author} {\bibfnamefont {J.}~\bibnamefont
  {Mondal}}, \bibinfo {author} {\bibfnamefont {N.}~\bibnamefont {Ahalawat}},
  \bibinfo {author} {\bibfnamefont {S.}~\bibnamefont {Pandit}}, \bibinfo
  {author} {\bibfnamefont {L.~E.}\ \bibnamefont {Kay}}, \ and\ \bibinfo
  {author} {\bibfnamefont {P.}~\bibnamefont {Vallurupalli}},\ }\href {\doibase
  10.1371/journal.pcbi.1006180} {\bibfield  {journal} {\bibinfo  {journal}
  {PLOS Computational Biology}\ }\textbf {\bibinfo {volume} {14}},\ \bibinfo
  {pages} {1} (\bibinfo {year} {2018})}\BibitemShut {NoStop}%
\bibitem [{\citenamefont {Ahalawat}\ and\ \citenamefont
  {Mondal}(2018{\natexlab{a}})}]{mondal_P450}%
  \BibitemOpen
  \bibfield  {author} {\bibinfo {author} {\bibfnamefont {N.}~\bibnamefont
  {Ahalawat}}\ and\ \bibinfo {author} {\bibfnamefont {J.}~\bibnamefont
  {Mondal}},\ }\href {\doibase 10.1021/jacs.8b10840} {\bibfield  {journal}
  {\bibinfo  {journal} {Journal of the American Chemical Society}\ }\textbf
  {\bibinfo {volume} {140}},\ \bibinfo {pages} {17743} (\bibinfo {year}
  {2018}{\natexlab{a}})},\ \Eprint
  {http://arxiv.org/abs/https://doi.org/10.1021/jacs.8b10840}
  {https://doi.org/10.1021/jacs.8b10840} \BibitemShut {NoStop}%
\bibitem [{\citenamefont {Plattner}\ and\ \citenamefont
  {Noe}(2015)}]{noe_trypsin}%
  \BibitemOpen
  \bibfield  {author} {\bibinfo {author} {\bibfnamefont {N.}~\bibnamefont
  {Plattner}}\ and\ \bibinfo {author} {\bibfnamefont {F.}~\bibnamefont {Noe}},\
  }\href@noop {} {\bibfield  {journal} {\bibinfo  {journal} {Nat Commun}\
  }\textbf {\bibinfo {volume} {6}},\ \bibinfo {pages} {7653} (\bibinfo {year}
  {2015})}\BibitemShut {NoStop}%
\bibitem [{\citenamefont {Tiwary}, \citenamefont {Mondal},\ and\ \citenamefont
  {Berne}(2017)}]{tiwary_dasatinib}%
  \BibitemOpen
  \bibfield  {author} {\bibinfo {author} {\bibfnamefont {P.}~\bibnamefont
  {Tiwary}}, \bibinfo {author} {\bibfnamefont {J.}~\bibnamefont {Mondal}}, \
  and\ \bibinfo {author} {\bibfnamefont {B.~J.}\ \bibnamefont {Berne}},\ }\href
  {\doibase 10.1126/sciadv.1700014} {\bibfield  {journal} {\bibinfo  {journal}
  {Science Advances}\ }\textbf {\bibinfo {volume} {3}} (\bibinfo {year}
  {2017}),\ 10.1126/sciadv.1700014},\ \Eprint
  {http://arxiv.org/abs/https://advances.sciencemag.org/content/3/5/e1700014.full.pdf}
  {https://advances.sciencemag.org/content/3/5/e1700014.full.pdf} \BibitemShut
  {NoStop}%
\bibitem [{\citenamefont {Buch}, \citenamefont {Giorgino},\ and\ \citenamefont
  {De~Fabritiis}(2011)}]{defabritiis}%
  \BibitemOpen
  \bibfield  {author} {\bibinfo {author} {\bibfnamefont {I.}~\bibnamefont
  {Buch}}, \bibinfo {author} {\bibfnamefont {T.}~\bibnamefont {Giorgino}}, \
  and\ \bibinfo {author} {\bibfnamefont {G.}~\bibnamefont {De~Fabritiis}},\
  }\href@noop {} {\bibfield  {journal} {\bibinfo  {journal} {Proc Natl Acad
  Sci}\ }\textbf {\bibinfo {volume} {108}},\ \bibinfo {pages} {10184} (\bibinfo
  {year} {2011})}\BibitemShut {NoStop}%
\bibitem [{\citenamefont {Mondal}, \citenamefont {Friesner},\ and\
  \citenamefont {Berne}(2014)}]{mondal_dasatinib}%
  \BibitemOpen
  \bibfield  {author} {\bibinfo {author} {\bibfnamefont {J.}~\bibnamefont
  {Mondal}}, \bibinfo {author} {\bibfnamefont {R.~A.}\ \bibnamefont
  {Friesner}}, \ and\ \bibinfo {author} {\bibfnamefont {B.~J.}\ \bibnamefont
  {Berne}},\ }\href@noop {} {\bibfield  {journal} {\bibinfo  {journal} {Journal
  of chemical theory and computation}\ }\textbf {\bibinfo {volume} {10}},\
  \bibinfo {pages} {5696} (\bibinfo {year} {2014})}\BibitemShut {NoStop}%
\bibitem [{\citenamefont {Chandler}(2005)}]{chandler}%
  \BibitemOpen
  \bibfield  {author} {\bibinfo {author} {\bibfnamefont {D.}~\bibnamefont
  {Chandler}},\ }\href@noop {} {\bibfield  {journal} {\bibinfo  {journal}
  {Nature}\ }\textbf {\bibinfo {volume} {437}},\ \bibinfo {pages} {640}
  (\bibinfo {year} {2005})}\BibitemShut {NoStop}%
\bibitem [{\citenamefont {Berne}, \citenamefont {Weeks},\ and\ \citenamefont
  {Zhou}(2009)}]{berneweekszhou}%
  \BibitemOpen
  \bibfield  {author} {\bibinfo {author} {\bibfnamefont {B.~J.}\ \bibnamefont
  {Berne}}, \bibinfo {author} {\bibfnamefont {J.~D.}\ \bibnamefont {Weeks}}, \
  and\ \bibinfo {author} {\bibfnamefont {R.}~\bibnamefont {Zhou}},\ }\href@noop
  {} {\bibfield  {journal} {\bibinfo  {journal} {Annual review of physical
  chemistry}\ }\textbf {\bibinfo {volume} {60}},\ \bibinfo {pages} {85}
  (\bibinfo {year} {2009})}\BibitemShut {NoStop}%
\bibitem [{\citenamefont {Rasaiah}, \citenamefont {Garde},\ and\ \citenamefont
  {Hummer}(2008)}]{hummer}%
  \BibitemOpen
  \bibfield  {author} {\bibinfo {author} {\bibfnamefont {J.~C.}\ \bibnamefont
  {Rasaiah}}, \bibinfo {author} {\bibfnamefont {S.}~\bibnamefont {Garde}}, \
  and\ \bibinfo {author} {\bibfnamefont {G.}~\bibnamefont {Hummer}},\
  }\href@noop {} {\bibfield  {journal} {\bibinfo  {journal} {Annual Review of
  Physical Chemistry}\ }\textbf {\bibinfo {volume} {59}},\ \bibinfo {pages}
  {713} (\bibinfo {year} {2008})}\BibitemShut {NoStop}%
\bibitem [{\citenamefont {Young}\ \emph {et~al.}(2007)\citenamefont {Young},
  \citenamefont {Abel}, \citenamefont {Kim}, \citenamefont {Berne},\ and\
  \citenamefont {Friesner}}]{young2007motifs}%
  \BibitemOpen
  \bibfield  {author} {\bibinfo {author} {\bibfnamefont {T.}~\bibnamefont
  {Young}}, \bibinfo {author} {\bibfnamefont {R.}~\bibnamefont {Abel}},
  \bibinfo {author} {\bibfnamefont {B.}~\bibnamefont {Kim}}, \bibinfo {author}
  {\bibfnamefont {B.~J.}\ \bibnamefont {Berne}}, \ and\ \bibinfo {author}
  {\bibfnamefont {R.~A.}\ \bibnamefont {Friesner}},\ }\href@noop {} {\bibfield
  {journal} {\bibinfo  {journal} {Proceedings of the National Academy of
  Sciences}\ }\textbf {\bibinfo {volume} {104}},\ \bibinfo {pages} {808}
  (\bibinfo {year} {2007})}\BibitemShut {NoStop}%
\bibitem [{\citenamefont {Braaten}, \citenamefont {Ansari},\ and\ \citenamefont
  {Luban}(1997)}]{hiv}%
  \BibitemOpen
  \bibfield  {author} {\bibinfo {author} {\bibfnamefont {D.}~\bibnamefont
  {Braaten}}, \bibinfo {author} {\bibfnamefont {H.}~\bibnamefont {Ansari}}, \
  and\ \bibinfo {author} {\bibfnamefont {J.}~\bibnamefont {Luban}},\
  }\href@noop {} {\bibfield  {journal} {\bibinfo  {journal} {Journal of
  virology}\ }\textbf {\bibinfo {volume} {71}},\ \bibinfo {pages} {2107}
  (\bibinfo {year} {1997})}\BibitemShut {NoStop}%
\bibitem [{\citenamefont {Mondal}, \citenamefont {Morrone},\ and\ \citenamefont
  {Berne}(2013)}]{mondal_fuller}%
  \BibitemOpen
  \bibfield  {author} {\bibinfo {author} {\bibfnamefont {J.}~\bibnamefont
  {Mondal}}, \bibinfo {author} {\bibfnamefont {J.~A.}\ \bibnamefont {Morrone}},
  \ and\ \bibinfo {author} {\bibfnamefont {B.~J.}\ \bibnamefont {Berne}},\
  }\href@noop {} {\bibfield  {journal} {\bibinfo  {journal} {Proceedings of the
  National Academy of Sciences}\ }\textbf {\bibinfo {volume} {110}},\ \bibinfo
  {pages} {13277} (\bibinfo {year} {2013})}\BibitemShut {NoStop}%
\bibitem [{\citenamefont {Tiwary}\ \emph {et~al.}(2015)\citenamefont {Tiwary},
  \citenamefont {Mondal}, \citenamefont {Morrone},\ and\ \citenamefont
  {Berne}}]{tiwary2015}%
  \BibitemOpen
  \bibfield  {author} {\bibinfo {author} {\bibfnamefont {P.}~\bibnamefont
  {Tiwary}}, \bibinfo {author} {\bibfnamefont {J.}~\bibnamefont {Mondal}},
  \bibinfo {author} {\bibfnamefont {J.~A.}\ \bibnamefont {Morrone}}, \ and\
  \bibinfo {author} {\bibfnamefont {B.}~\bibnamefont {Berne}},\ }\href@noop {}
  {\bibfield  {journal} {\bibinfo  {journal} {Proc. Natl. Acad. Sci.}\ }\textbf
  {\bibinfo {volume} {112}},\ \bibinfo {pages} {12015} (\bibinfo {year}
  {2015})}\BibitemShut {NoStop}%
\bibitem [{\citenamefont {Tiwary}\ and\ \citenamefont
  {Berne}(2016{\natexlab{a}})}]{tiwary_wet}%
  \BibitemOpen
  \bibfield  {author} {\bibinfo {author} {\bibfnamefont {P.}~\bibnamefont
  {Tiwary}}\ and\ \bibinfo {author} {\bibfnamefont {B.~J.}\ \bibnamefont
  {Berne}},\ }\href {\doibase 10.1063/1.4959969} {\bibfield  {journal}
  {\bibinfo  {journal} {The Journal of Chemical Physics}\ }\textbf {\bibinfo
  {volume} {145}},\ \bibinfo {pages} {054113} (\bibinfo {year}
  {2016}{\natexlab{a}})},\ \Eprint
  {http://arxiv.org/abs/https://doi.org/10.1063/1.4959969}
  {https://doi.org/10.1063/1.4959969} \BibitemShut {NoStop}%
\bibitem [{\citenamefont {Setny}, \citenamefont {Baron},\ and\ \citenamefont
  {McCammon}(2010)}]{setny_jctc}%
  \BibitemOpen
  \bibfield  {author} {\bibinfo {author} {\bibfnamefont {P.}~\bibnamefont
  {Setny}}, \bibinfo {author} {\bibfnamefont {R.}~\bibnamefont {Baron}}, \ and\
  \bibinfo {author} {\bibfnamefont {J.~A.}\ \bibnamefont {McCammon}},\ }\href
  {\doibase 10.1021/ct1003077} {\bibfield  {journal} {\bibinfo  {journal}
  {Journal of Chemical Theory and Computation}\ }\textbf {\bibinfo {volume}
  {6}},\ \bibinfo {pages} {2866} (\bibinfo {year} {2010})},\ \bibinfo {note}
  {pMID: 20844599},\ \Eprint
  {http://arxiv.org/abs/https://doi.org/10.1021/ct1003077}
  {https://doi.org/10.1021/ct1003077} \BibitemShut {NoStop}%
\bibitem [{\citenamefont {Baron}, \citenamefont {Setny},\ and\ \citenamefont
  {Andrew~McCammon}(2010)}]{mccammon_jacs}%
  \BibitemOpen
  \bibfield  {author} {\bibinfo {author} {\bibfnamefont {R.}~\bibnamefont
  {Baron}}, \bibinfo {author} {\bibfnamefont {P.}~\bibnamefont {Setny}}, \ and\
  \bibinfo {author} {\bibfnamefont {J.}~\bibnamefont {Andrew~McCammon}},\
  }\href {\doibase 10.1021/ja1050082} {\bibfield  {journal} {\bibinfo
  {journal} {Journal of the American Chemical Society}\ }\textbf {\bibinfo
  {volume} {132}},\ \bibinfo {pages} {12091} (\bibinfo {year} {2010})},\
  \bibinfo {note} {pMID: 20695475},\ \Eprint
  {http://arxiv.org/abs/http://dx.doi.org/10.1021/ja1050082}
  {http://dx.doi.org/10.1021/ja1050082} \BibitemShut {NoStop}%
\bibitem [{\citenamefont {Setny}\ \emph {et~al.}(2013)\citenamefont {Setny},
  \citenamefont {Baron}, \citenamefont {Michael Kekenes-Huskey}, \citenamefont
  {McCammon},\ and\ \citenamefont {Dzubiella}}]{setny_pnas}%
  \BibitemOpen
  \bibfield  {author} {\bibinfo {author} {\bibfnamefont {P.}~\bibnamefont
  {Setny}}, \bibinfo {author} {\bibfnamefont {R.}~\bibnamefont {Baron}},
  \bibinfo {author} {\bibfnamefont {P.}~\bibnamefont {Michael Kekenes-Huskey}},
  \bibinfo {author} {\bibfnamefont {J.~A.}\ \bibnamefont {McCammon}}, \ and\
  \bibinfo {author} {\bibfnamefont {J.}~\bibnamefont {Dzubiella}},\ }\href
  {\doibase 10.1073/pnas.1221231110} {\bibfield  {journal} {\bibinfo  {journal}
  {Proceedings of the National Academy of Sciences}\ }\textbf {\bibinfo
  {volume} {110}},\ \bibinfo {pages} {1197} (\bibinfo {year} {2013})},\ \Eprint
  {http://arxiv.org/abs/https://www.pnas.org/content/110/4/1197.full.pdf}
  {https://www.pnas.org/content/110/4/1197.full.pdf} \BibitemShut {NoStop}%
\bibitem [{\citenamefont {Setny}\ and\ \citenamefont
  {Geller}(2006)}]{setny_jcp}%
  \BibitemOpen
  \bibfield  {author} {\bibinfo {author} {\bibfnamefont {P.}~\bibnamefont
  {Setny}}\ and\ \bibinfo {author} {\bibfnamefont {M.}~\bibnamefont {Geller}},\
  }\href {\doibase 10.1063/1.2355487} {\bibfield  {journal} {\bibinfo
  {journal} {The Journal of Chemical Physics}\ }\textbf {\bibinfo {volume}
  {125}},\ \bibinfo {pages} {144717} (\bibinfo {year} {2006})},\ \Eprint
  {http://arxiv.org/abs/https://doi.org/10.1063/1.2355487}
  {https://doi.org/10.1063/1.2355487} \BibitemShut {NoStop}%
\bibitem [{\citenamefont {Hernandez}\ \emph {et~al.}(2013)\citenamefont
  {Hernandez}, \citenamefont {Paul}, \citenamefont {Giorgino}, \citenamefont
  {Fabritiis},\ and\ \citenamefont {Noe}}]{tica1}%
  \BibitemOpen
  \bibfield  {author} {\bibinfo {author} {\bibfnamefont {G.~P.}\ \bibnamefont
  {Hernandez}}, \bibinfo {author} {\bibfnamefont {F.}~\bibnamefont {Paul}},
  \bibinfo {author} {\bibfnamefont {T.}~\bibnamefont {Giorgino}}, \bibinfo
  {author} {\bibfnamefont {G.~D.}\ \bibnamefont {Fabritiis}}, \ and\ \bibinfo
  {author} {\bibfnamefont {F.}~\bibnamefont {Noe}},\ }\href {\doibase
  10.1063/1.4811489} {\bibfield  {journal} {\bibinfo  {journal} {The Journal of
  Chemical Physics}\ }\textbf {\bibinfo {volume} {139}},\ \bibinfo {pages}
  {015102} (\bibinfo {year} {2013})}\BibitemShut {NoStop}%
\bibitem [{\citenamefont {Schwantes}\ and\ \citenamefont
  {Pande}(2013)}]{tica5}%
  \BibitemOpen
  \bibfield  {author} {\bibinfo {author} {\bibfnamefont {C.~R.}\ \bibnamefont
  {Schwantes}}\ and\ \bibinfo {author} {\bibfnamefont {V.~S.}\ \bibnamefont
  {Pande}},\ }\href {\doibase 10.1021/ct300878a} {\bibfield  {journal}
  {\bibinfo  {journal} {Journal of Chemical Theory and Computation}\ }\textbf
  {\bibinfo {volume} {9}},\ \bibinfo {pages} {2000} (\bibinfo {year}
  {2013})}\BibitemShut {NoStop}%
\bibitem [{\citenamefont {Husic}\ and\ \citenamefont
  {Pande}(2018)}]{msm_review}%
  \BibitemOpen
  \bibfield  {author} {\bibinfo {author} {\bibfnamefont {B.~E.}\ \bibnamefont
  {Husic}}\ and\ \bibinfo {author} {\bibfnamefont {V.~S.}\ \bibnamefont
  {Pande}},\ }\href {\doibase 10.1021/jacs.7b12191} {\bibfield  {journal}
  {\bibinfo  {journal} {Journal of the American Chemical Society}\ }\textbf
  {\bibinfo {volume} {140}},\ \bibinfo {pages} {2386} (\bibinfo {year}
  {2018})},\ \bibinfo {note} {pMID: 29323881},\ \Eprint
  {http://arxiv.org/abs/https://doi.org/10.1021/jacs.7b12191}
  {https://doi.org/10.1021/jacs.7b12191} \BibitemShut {NoStop}%
\bibitem [{\citenamefont {Jorgensen}\ \emph {et~al.}(1983)\citenamefont
  {Jorgensen}, \citenamefont {Chandrasekhar}, \citenamefont {Madura},
  \citenamefont {Impey},\ and\ \citenamefont {Klein}}]{tip4p}%
  \BibitemOpen
  \bibfield  {author} {\bibinfo {author} {\bibfnamefont {W.~L.}\ \bibnamefont
  {Jorgensen}}, \bibinfo {author} {\bibfnamefont {J.}~\bibnamefont
  {Chandrasekhar}}, \bibinfo {author} {\bibfnamefont {J.~D.}\ \bibnamefont
  {Madura}}, \bibinfo {author} {\bibfnamefont {R.~W.}\ \bibnamefont {Impey}}, \
  and\ \bibinfo {author} {\bibfnamefont {M.~L.}\ \bibnamefont {Klein}},\
  }\href@noop {} {\bibfield  {journal} {\bibinfo  {journal} {The Journal of
  chemical physics}\ }\textbf {\bibinfo {volume} {79}},\ \bibinfo {pages} {926}
  (\bibinfo {year} {1983})}\BibitemShut {NoStop}%
\bibitem [{\citenamefont {Bussi}, \citenamefont {Donadio},\ and\ \citenamefont
  {Parrinello}(2007)}]{bussidonadio}%
  \BibitemOpen
  \bibfield  {author} {\bibinfo {author} {\bibfnamefont {G.}~\bibnamefont
  {Bussi}}, \bibinfo {author} {\bibfnamefont {D.}~\bibnamefont {Donadio}}, \
  and\ \bibinfo {author} {\bibfnamefont {M.}~\bibnamefont {Parrinello}},\
  }\href@noop {} {\bibfield  {journal} {\bibinfo  {journal} {The Journal of
  chemical physics}\ }\textbf {\bibinfo {volume} {126}},\ \bibinfo {pages}
  {014101} (\bibinfo {year} {2007})}\BibitemShut {NoStop}%
\bibitem [{\citenamefont {Darden}, \citenamefont {York},\ and\ \citenamefont
  {Pederson}(1993)}]{pme1}%
  \BibitemOpen
  \bibfield  {author} {\bibinfo {author} {\bibfnamefont {T.}~\bibnamefont
  {Darden}}, \bibinfo {author} {\bibfnamefont {D.}~\bibnamefont {York}}, \ and\
  \bibinfo {author} {\bibfnamefont {L.~G.}\ \bibnamefont {Pederson}},\
  }\href@noop {} {\bibfield  {journal} {\bibinfo  {journal} {J. Chem. Phys.}\
  }\textbf {\bibinfo {volume} {98}},\ \bibinfo {pages} {952} (\bibinfo {year}
  {1993})}\BibitemShut {NoStop}%
\bibitem [{\citenamefont {Scherer}\ \emph {et~al.}(2015)\citenamefont
  {Scherer}, \citenamefont {Trendelkamp-Schroer}, \citenamefont {Paul},
  \citenamefont {Perez-Hernandez}, \citenamefont {Hoffmann}, \citenamefont
  {Plattner}, \citenamefont {Wehmeyer}, \citenamefont {Prinz},\ and\
  \citenamefont {Noe}}]{pyemma}%
  \BibitemOpen
  \bibfield  {author} {\bibinfo {author} {\bibfnamefont {M.~K.}\ \bibnamefont
  {Scherer}}, \bibinfo {author} {\bibfnamefont {B.}~\bibnamefont
  {Trendelkamp-Schroer}}, \bibinfo {author} {\bibfnamefont {F.}~\bibnamefont
  {Paul}}, \bibinfo {author} {\bibfnamefont {G.}~\bibnamefont
  {Perez-Hernandez}}, \bibinfo {author} {\bibfnamefont {M.}~\bibnamefont
  {Hoffmann}}, \bibinfo {author} {\bibfnamefont {N.}~\bibnamefont {Plattner}},
  \bibinfo {author} {\bibfnamefont {C.}~\bibnamefont {Wehmeyer}}, \bibinfo
  {author} {\bibfnamefont {J.-H.}\ \bibnamefont {Prinz}}, \ and\ \bibinfo
  {author} {\bibfnamefont {F.}~\bibnamefont {Noe}},\ }\href {\doibase
  10.1021/acs.jctc.5b00743} {\bibfield  {journal} {\bibinfo  {journal} {Journal
  of Chemical Theory and Computation}\ }\textbf {\bibinfo {volume} {11}},\
  \bibinfo {pages} {5525} (\bibinfo {year} {2015})}\BibitemShut {NoStop}%
\bibitem [{\citenamefont {Wehmeyer}\ \emph {et~al.}(2018)\citenamefont
  {Wehmeyer}, \citenamefont {Scherer}, \citenamefont {Hempel}, \citenamefont
  {Husic}, \citenamefont {Olsson},\ and\ \citenamefont
  {Noé}}]{pyemma_tutorial}%
  \BibitemOpen
  \bibfield  {author} {\bibinfo {author} {\bibfnamefont {C.}~\bibnamefont
  {Wehmeyer}}, \bibinfo {author} {\bibfnamefont {M.~K.}\ \bibnamefont
  {Scherer}}, \bibinfo {author} {\bibfnamefont {T.}~\bibnamefont {Hempel}},
  \bibinfo {author} {\bibfnamefont {B.~E.}\ \bibnamefont {Husic}}, \bibinfo
  {author} {\bibfnamefont {S.}~\bibnamefont {Olsson}}, \ and\ \bibinfo {author}
  {\bibfnamefont {F.}~\bibnamefont {Noé}},\ }\href {\doibase
  10.33011/livecoms.1.1.5965} {\bibfield  {journal} {\bibinfo  {journal}
  {Living Journal of Computational Molecular Science}\ }\textbf {\bibinfo
  {volume} {1}},\ \bibinfo {pages} {5965} (\bibinfo {year} {2018})}\BibitemShut
  {NoStop}%
\bibitem [{\citenamefont {Lloyd}(2006)}]{kmeans}%
  \BibitemOpen
  \bibfield  {author} {\bibinfo {author} {\bibfnamefont {S.}~\bibnamefont
  {Lloyd}},\ }\href {\doibase 10.1109/TIT.1982.1056489} {\bibfield  {journal}
  {\bibinfo  {journal} {IEEE Trans. Inf. Theor.}\ }\textbf {\bibinfo {volume}
  {28}},\ \bibinfo {pages} {129} (\bibinfo {year} {2006})}\BibitemShut
  {NoStop}%
\bibitem [{pye()}]{pyemma_github}%
  \BibitemOpen
  \href@noop {} {\enquote {\bibinfo {title} {pyemma tutorial github},}\
  }\bibinfo {howpublished}
  {\url{https://github.com/markovmodel/pyemma_tutorials/blob/master/notebooks/06-expectations-and-observables.ipynb}},\
  \bibinfo {note} {accessed: 2019-09-30}\BibitemShut {NoStop}%
\bibitem [{\citenamefont {Ahalawat}\ and\ \citenamefont
  {Mondal}(2018{\natexlab{b}})}]{mondal_gb1}%
  \BibitemOpen
  \bibfield  {author} {\bibinfo {author} {\bibfnamefont {N.}~\bibnamefont
  {Ahalawat}}\ and\ \bibinfo {author} {\bibfnamefont {J.}~\bibnamefont
  {Mondal}},\ }\href {\doibase 10.1063/1.5041073} {\bibfield  {journal}
  {\bibinfo  {journal} {The Journal of Chemical Physics}\ }\textbf {\bibinfo
  {volume} {149}},\ \bibinfo {pages} {094101} (\bibinfo {year}
  {2018}{\natexlab{b}})},\ \Eprint
  {http://arxiv.org/abs/https://doi.org/10.1063/1.5041073}
  {https://doi.org/10.1063/1.5041073} \BibitemShut {NoStop}%
\bibitem [{\citenamefont {E.}\ and\ \citenamefont
  {Vanden-Eijnden}(2006)}]{tpt1}%
  \BibitemOpen
  \bibfield  {author} {\bibinfo {author} {\bibfnamefont {W.}~\bibnamefont
  {E.}}\ and\ \bibinfo {author} {\bibfnamefont {E.}~\bibnamefont
  {Vanden-Eijnden}},\ }\href {\doibase 10.1007/s10955-005-9003-9} {\bibfield
  {journal} {\bibinfo  {journal} {Journal of Statistical Physics}\ }\textbf
  {\bibinfo {volume} {123}},\ \bibinfo {pages} {503} (\bibinfo {year}
  {2006})}\BibitemShut {NoStop}%
\bibitem [{\citenamefont {No{\'e}}\ \emph {et~al.}(2009)\citenamefont
  {No{\'e}}, \citenamefont {Sch{\"u}tte}, \citenamefont {Vanden-Eijnden},
  \citenamefont {Reich},\ and\ \citenamefont {Weikl}}]{tpt3}%
  \BibitemOpen
  \bibfield  {author} {\bibinfo {author} {\bibfnamefont {F.}~\bibnamefont
  {No{\'e}}}, \bibinfo {author} {\bibfnamefont {C.}~\bibnamefont
  {Sch{\"u}tte}}, \bibinfo {author} {\bibfnamefont {E.}~\bibnamefont
  {Vanden-Eijnden}}, \bibinfo {author} {\bibfnamefont {L.}~\bibnamefont
  {Reich}}, \ and\ \bibinfo {author} {\bibfnamefont {T.~R.}\ \bibnamefont
  {Weikl}},\ }\href {\doibase 10.1073/pnas.0905466106} {\bibfield  {journal}
  {\bibinfo  {journal} {Proceedings of the National Academy of Sciences}\
  }\textbf {\bibinfo {volume} {106}},\ \bibinfo {pages} {19011} (\bibinfo
  {year} {2009})},\ \Eprint
  {http://arxiv.org/abs/http://www.pnas.org/content/106/45/19011.full.pdf}
  {http://www.pnas.org/content/106/45/19011.full.pdf} \BibitemShut {NoStop}%
\bibitem [{\citenamefont {Tiwary}\ and\ \citenamefont
  {Berne}(2016{\natexlab{b}})}]{sgoop}%
  \BibitemOpen
  \bibfield  {author} {\bibinfo {author} {\bibfnamefont {P.}~\bibnamefont
  {Tiwary}}\ and\ \bibinfo {author} {\bibfnamefont {B.~J.}\ \bibnamefont
  {Berne}},\ }\href {\doibase 10.1073/pnas.1600917113} {\bibfield  {journal}
  {\bibinfo  {journal} {Proceedings of the National Academy of Sciences}\
  }\textbf {\bibinfo {volume} {113}},\ \bibinfo {pages} {2839} (\bibinfo {year}
  {2016}{\natexlab{b}})},\ \Eprint
  {http://arxiv.org/abs/https://www.pnas.org/content/113/11/2839.full.pdf}
  {https://www.pnas.org/content/113/11/2839.full.pdf} \BibitemShut {NoStop}%
\end{thebibliography}

%
\end{document}